3-2016

# An Artificial Neural Networks based Temperature Prediction Framework for Network-on-Chip based Multicore Platform

Sandeep Aswath Narayana
sa5641@rit.edu

# An Artificial Neural Networks based Temperature Prediction Framework for Network-on-Chip based Multicore Platform

By
**Sandeep Aswath Narayana**

A Thesis Submitted in Partial Fulfillment of the Requirements for the Degree of
**Master of Science in Electrical Engineering**

Supervised by
**Dr. Amlan Ganguly**

Department of Computer Engineering
Kate Gleason College of Engineering
Rochester Institute of Technology
Rochester, NY
March 2016

**Approved By:**

___

Dr. Amlan Ganguly
*Thesis Advisor – R.I.T. Dept. of Computer Engineering*

___

Dr. Ray Ptucha
*Secondary Advisor – R.I.T. Dept. of Computer Engineering*

___

Dr. Mehran Mozaffari Kermani
*Secondary Advisor – R.I.T. Dept. of Electrical and Microelectronic Engineering*

___

Dr. Sohail A. Dianat
*Professor Department Head, R.I.T. Dept. of Electrical and Microelectronic Engineering*

*To my beloved parents Mr. Aswath Narayana and Mrs. Shubha, and my precious brother Aryan Rajeev.*

# Acknowledgements

I take this opportunity to express my profound gratitude and deep regards to my primary advisor Dr. Amlan Ganguly for his exemplary guidance, monitoring and constant encouragement throughout this thesis. Dr. Amlan dedicated his valuable time to review my work constantly and provide valuable suggestions which helped in overcoming many obstacles and keeping the work on the right track. I would like to express my deepest gratitude to Dr. Ray Ptucha and Dr. Mehran Mozaffari Kermani for sharing their thoughts and suggesting valuable ideas which have had significant impact on this thesis. I am grateful for their valuable time and cooperation during the course of this thesis. I also take this opportunity to thank my research group members for all the constant support and help provided by them.

# Abstract

Continuous improvement in silicon process technologies has made possible the integration of hundreds of cores on a single chip. However, power and heat have become dominant constraints in designing these massive multicore chips causing issues with reliability, timing variations and reduced lifetime of the chips. Dynamic Thermal Management (DTM) is a solution to avoid high temperatures on the die. Typical DTM schemes only address core level thermal issues. However, the Network-on-chip (NoC) paradigm, which has emerged as an enabling methodology for integrating hundreds to thousands of cores on the same die can contribute significantly to the thermal issues. Moreover, the typical DTM is triggered reactively based on temperature measurements from on-chip thermal sensor requiring long reaction times whereas predictive DTM method estimates future temperature in advance, eliminating the chance of temperature overshoot. Artificial Neural Networks (ANNs) have been used in various domains for modeling and prediction with high accuracy due to its ability to learn and adapt. This thesis concentrates on designing an ANN prediction engine to predict the thermal profile of the cores and Network-on-Chip elements of the chip. This thermal profile of the chip is then used by the predictive DTM that combines both core level and network level DTM techniques. On-chip wireless interconnect which is recently envisioned to enable energy-efficient data exchange between cores in a multicore environment, will be used to provide a broadcast-capable medium to efficiently distribute thermal control messages to trigger and manage the DTM schemes.

# Table of Contents

# List of Figures

# List of Tables

# Chapter 1. Introduction

Modern computing systems are becoming exceptionally complex to keep up with increasing performance demands of applications. The number of cores on a die are continuously increasing and the transistor counts have reached billions. This scaling causes unprecedented increase in thermal problems. This in turn leads to thermal variation and reliability issues. Moreover, prolonged operation at high temperatures reduces lifetime of the chip. Dynamic Thermal Management (DTM) is an active research area that uses mechanisms like Task Reallocation, Dynamic Voltage Frequency Scaling (DVFS), Clock/Power Gating to mitigate these thermal issues. These DTM techniques are already being used in state of the art multicore processors like Single Chip Cloud Computer and Intel Core I7 processors [1]. However, the Network-on-chip (NoC) paradigm, which has emerged as an enabling methodology for integrating hundreds to thousands of cores on the same die [7], can contribute significantly to these thermal issues. Typical DTM techniques address either core level or NoC level thermal problems. However, an effective DTM technique should consider both core level and NoC level thermal management scheme. Moreover, it should be pro-active instead of reactive as in reactive DTM techniques have long response times, which can lead to transient temperature overshoots. In proactive DTM methods, hotspots are predicted before they are actually created eliminating the use of thermal sensors. In [4], authors propose a Look-Up Table (LUT) based thermal prediction method. However, the large memory requirements to store the LUT for large system sizes and the associated computational overhead make it a non-scalable mechanism. On the other hand, Artificial Neural Networks (ANNs) have proven to be highly accurate to learn and adapt to a pattern which is used to modeling and prediction [5]. In [6], an ANN based predictor is proposed to monitor inter-core traffic congestion in a multicore chip. This thesis proposes to design an ANN based prediction engine to predict the thermal characteristics of the chip elements. The proposed ANN engine is used to trigger the DTM scheme that combines both core level and NoC level thermal management technique. Starting from Network-on-chip paradigm, this chapter discusses DTM techniques and ANN prediction engine.

## 1.1 Network-on-Chip (NoC)

With the emergence of the multicore era, more and more cores within a single chip are being integrated to exploit core level parallelism. However, with increasing number of cores, the length of the wires connecting the cores keep increasing. This in turn results in high inter-core communication delay. Moreover, the global wires do not scale down with technology. Even with repeater insertion, the wire delay may exceed a single clock cycle. The Network-on-Chip (NoC) paradigm has emerged as an interconnection infrastructure to mitigate global wire delays by designing separate scalable interconnection fabrics to support high speed communication between cores. Generally, in NoC, data is transferred via wormhole switching using virtual channel (VC) based switches. Data packets are broken down and transferred in the form of flow control units or flits which are the smallest amount of information in a packet that can be transferred between adjacent switches in one clock cycle [34].

Traditional NoC architectures are based on planar metallic interconnect. However, due to multi-hop data communication over metal interconnects, these architectures suffer from latency and high energy consumption. The quest to have energy efficient NoC has resulted in exploring emerging interconnect technologies like photonics, 3D integration, RF and wireless interconnects for on-chip inter-core data transfers. Three-dimensional interconnects integrate multiple active layers onto a single chip and consequently reduces the hop-count and the average wire length of a single hop. The performance advantages of three dimensional interconnects come at the cost of an increase in temperatures due to smaller foot print and higher resultant power densities which cause high heat dissipation [44] and require sophisticated cooling mechanism[45]. Fabrication of 3D interconnections is also proven to be challenging due to the issues with inter-layer alignments, bonding, inter-layer contact patterning[17] and increased risks of manufacturing defects. The on-chip photonic interconnects are implemented using on-chip optical waveguides, micro-ring resonators and laser sources. With the data transmissions occurring at the speed of light, photonic interconnects have been predicted to considerably enhance the bandwidth and reduce latency [18]. Due to the low loss seen in the optical waveguides, regeneration or buffering is not required to ensure reliable data transmission through the photonic interconnects. However, the technological intricacy involved in manufacturing these photonic devices and integrating them with silicon-compatible circuits under area, power and delay constraints is a non-trivial challenge.

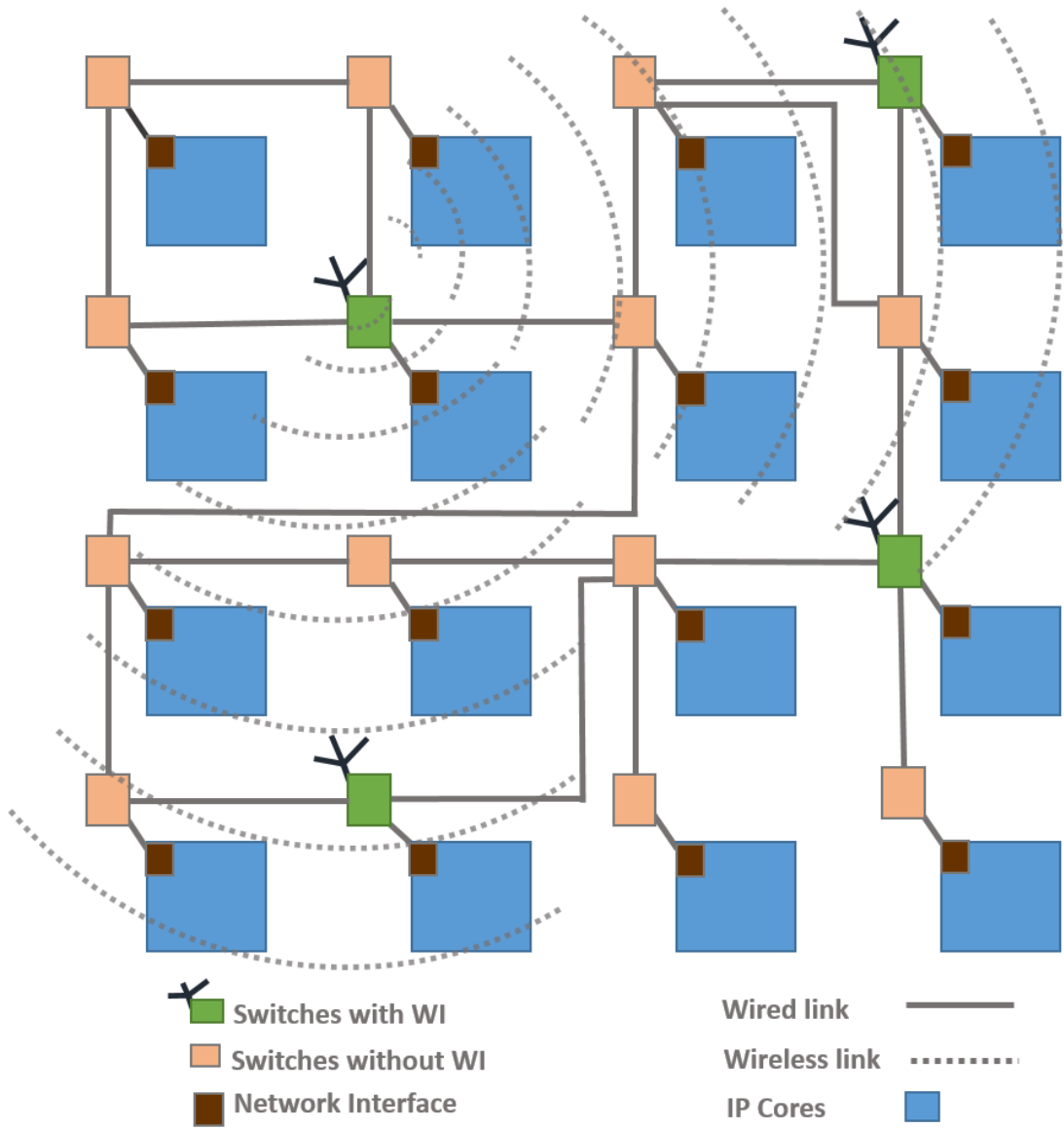

Fig. 1: Architecture for broadcast capable small-world WiNoC

Multiband RF interconnects use wires as transmission lines to transfer data in the form of electromagnetic (EM) waves. The data is modulated onto a carrier using amplitude or phase shift keying [43]. Using this method of interconnection, bandwidth of conventional wires can be increased using multiple access techniques resulting in low latency data transfer at speed of light via EM waves. Multiband RF interconnects are limited by the design of high frequency oscillators and filters on the chip for the transceivers.

On-chip wireless interconnect is a promising alternative to the performance limitations seen by long-distance wired links. On-chip wireless transceivers enable long distance, high bandwidth and low latency communication over long range paths. Absence of the need for physical interconnection layouts makes wireless interconnects stand out from other emerging interconnects. Several WiNoC architectures have been explored in recent literature. Small-world networks as shown in Fig. 1, is a type of complex networks often found in nature that are characterized by both short-distance and long-range links [30]. This improves the efficiency of the network as they have very low average number of hops between nodes even for very large network sizes. Hence, such network topologies are suitable for designing scalable, hybrid WiNoCs where conventional wireline interconnects are augmented with long-range wireless shortcuts. This thesis adopts the method outlined in [29] for the establishment of the wireless links overlaid on a wireline NoC architecture inspired from small-world network topologies

The antenna and the transceiver are the two principle components of the WiNoC architecture. The on-chip antenna for the WiNoC has to provide the best power gain for the smallest area overhead. Zig-zag antennas are non-directional and also demonstrate the characteristics of higher power gain for and smaller power gain [24]. These characteristics of the zig-zag antennas make it suitable for our application where the Wireless Interconnect (WI) attached to the Artificial Neural Network (ANN) needs to broadcast thermal control messages to all other WIs deployed in different parts of the chip. The antenna design from [9] provides a 3dB bandwidth of 16 GHz with a center frequency around 60GHz for a communication range of 20 mm which is adopted in this thesis. The quarter wave antenna uses an axial length of 0.38 mm in the silicon substrate for optimal power efficiency.

Advances in antenna and mm-wave transceiver design in standard bulk CMOS technologies have made on-chip wireless interconnect feasible. However, the limited bandwidth of the wireless channels at such high frequencies limits the achievable performance benefits. The mm-wave WIs inherent capability to share the wireless channel and provide a broadcast capability. This property of mm-wave has been shown to be beneficial for the exchange of control and synchronization information across a multicore chip. Designing wireless transceivers in multiple frequency bands for enhancing the performance of the NoC is a non-trivial challenge and is not scalable in the near future. One way of avoiding interference and contention between multiple transmitters is wireless token passing protocol to give access to the medium to a single transmitter at a time. The token passing scheme eliminates the need for centralized control and arbitration among the transceivers, which might be located in distant parts of the die. Hence, the token passing scheme has been adopted in multiple WiNoC designs. A single-bit register in the wireless switches can denote the presence of the token at a WI to minimize the associated hardware. When this register is set, it enables that particular WI to transmit data flits over the wireless medium. When the WI is done with its transmission it passes the token to the next WI in a round robin fashion. The token flit consists of two fields, nextWI and prevWI. The prevWI denotes the ID of the WI that released the token and the nextWI denotes the ID of the WI that will possess the token next. Each WI possesses the token for a maximum period of time before releasing it to the next WI.

## 1.2 Dynamic Thermal Management (DTM)

The continuous scaling of the chip dimensions with the integration of hundreds of cores and NoC components leads to uneven distribution of network traffic and workloads, causing formation hotspots. If thermal hotspots are not addressed initially, then this affects the neighboring NoC components or cores in a domino style resulting in timing variations, reliability, and eventually reduce the lifetime of the chip. Dynamic Thermal Management (DTM) being an active research area, is used to address the above mentioned thermal issues. Traditional DTM like Dynamic Frequency Scaling (DFS), Dynamic Voltage Scaling (DVS), clock gating, slowdown the CPU computation to reduce heat dissipation. Although they could effectively reduce temperature, they incur significant performance overhead. Task migration redistributes the existing processes based on the current thermal profile of the chip to reduce the peak temperature without throttling the computation. However, most of the existing task migration techniques are employed based on the temperature estimated from the thermal sensors. This makes the schemes reactive, in turn, requiring long reaction times. Moreover, the characteristics and reliability of the sensors affect the effectiveness of the scheme significantly [2]. Consequently proactive or predictive DTM mechanisms have received attention in recent times [3] [4]. In [4], a Look up Table (LUT) based thermal estimator is proposed. In such a predictor, the LUT characterizes the thermal response of the chip to the varying power dissipation profiles, which can be used to predict the future chip temperature. However, for highly scaled system sizes, the size of the LUT is very large and the associated memory requirements make it an unviable. ANNs have been known to be useful for prediction algorithms, and form a viable option for thermal estimation. In this thesis, we use an ANN based thermal estimator for reasons discussed in the following section.

## 1.3    Artificial Neural Network

The Artificial Neural Network (ANN) is a machine learning prediction algorithm that is inspired by animal nervous system. In recent literature ANN is used to predict the network traffic pattern in a NoC environment [6]. ANN is considered as it is not a complex algorithm for hardware implementation.

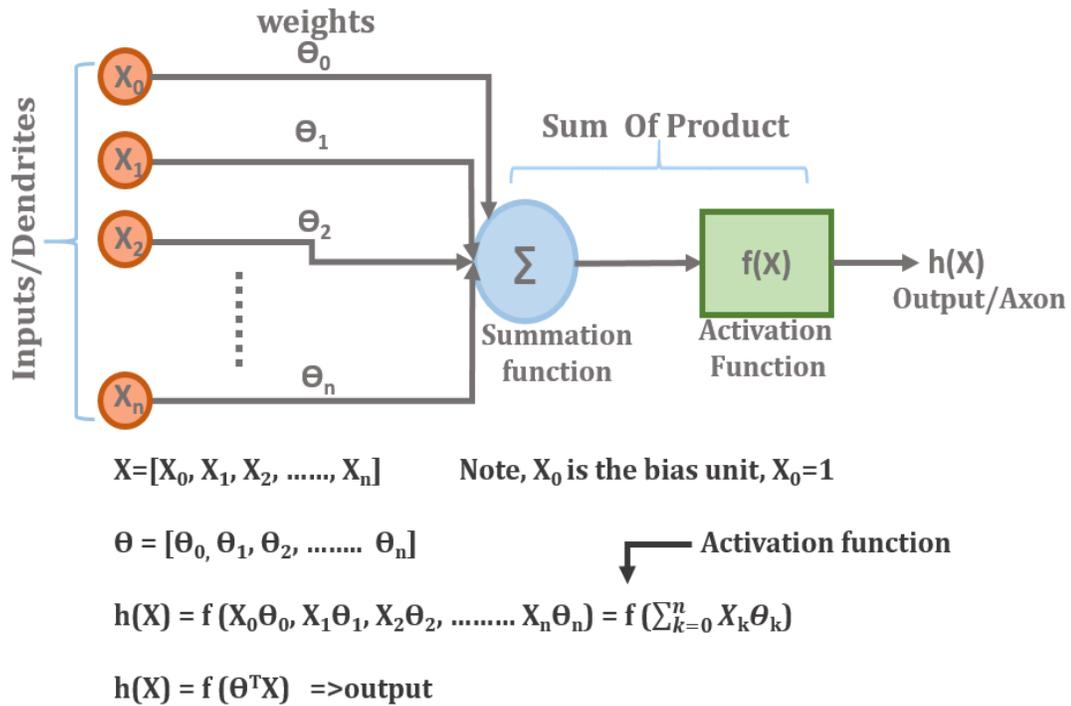

$X = [X_0, X_1, X_2, \ldots, X_n]$  Note, $X_0$ is the bias unit, $X_0 = 1$

$\theta = [\theta_0, \theta_1, \theta_2, \ldots \theta_n]$   — Activation function

$h(X) = f(X_0\theta_0, X_1\theta_1, X_2\theta_2, \ldots X_n\theta_n) = f\left(\sum_{k=0}^{n} X_k\theta_k\right)$

$h(X) = f(\theta^T X)$  =>output

Fig. 2: Mathematical Model of a single neuron.

First, to understand the overview of how an ANN works, consider a single neuron model as shown in the Fig.2. The dimension of the input sample is *n-1* as $X_0$ is the biased input. The weights '$\theta$' are initialized to small ± values centered on *0*. When the *n* dimensional input (including the bias input) is sent through the ANN it is initially multiplied by its initial weights. When the weighted sum of the n dimensions exceed a threshold value then the neuron fires a floating point value which is then passed through the activation function $f(\theta^T X)$ for mapping the value to give an output. When Cost Function is high, Backpropagation algorithm is used in changing the weights of the ANN to have a low cost. In the next cycle, next input is fed and simultaneously the weights are updated. The above process is recursively followed, until we have a low cost function. Multiple such neurons connected together is known as a Neural Network. The basic operation of processing the inputs to predict the output, make the ANN design easy to implement in VLSI. The detailed explanation of the ANN is given in the later section of the thesis.

## 1.4 Thesis Contribution

The following summarizes the contribution of this work:

1) **Proposed Artificial Neural Network to Predict Temperature profile in a Multi-core chip**

    a) Training dataset was collected from thermal Simulator and labeled for training of the ANN.

    b) Designed an optimized ANN network with optimal latency, area, and memory requirements.

    c) Evaluated the ANN on comparison look-up-table approaches

2) **Integration of ANN based Thermal predictor with Dynamic Thermal Management**

    a) Integrated the ANN based thermal predictor with Dynamic Thermal management scheme for a 64 core WiNoC based architecture.

3) **Evaluation of thermal characteristics and performance of the WiNoC with Dynamic Thermal Management utilizing the proposed ANN based thermal predicted.**

# Chapter 2. Related Work

The problem of solving thermal hotspots prevention using Dynamic Thermal Management (DTM) of multicore system is an active researched field. Task migration redistributes existing processes to available cores based on the current thermal profile of the chip. Runtime task migration is a popular technique for reducing peak temperature. There are numerous workload migration schemes and also distributed migration schemes that are proposed in [15]. Systems with dynamic technique, respond to real-time changes and adapt to the current workload. Considering heterogeneous and morphable cores, an efficient task-mapping algorithm under power constraints is proposed in [17]. Autoregressive moving average and lookup table based thermal predictor are used in the DTM [3]. The temperature of multicore system for a user-defined threshold is controlled using convex optimization [19]. Thermal Herd, proposed in [20], provides a distributed runtime scheme for thermal management that allows NoC routers to collaboratively regulate the network temperature profile and work to avert thermal emergencies while minimizing performance impact. This work primarily targeted mesh-based wireline NoCs.

Artificial neural networks (ANNs) have been used in various domains for modeling and prediction with high accuracy due to their ability to learn and adapt [5]. Recently, an ANN based predictor was designed to monitor and avoid inter-core traffic congestion in a multicore chip [6]. ANN has the adaptability to changing traffic conditions and their ability to learn about small network spatiotemporal variations which can lead to online congestion and thus build on improving their ability to forecast the next hotspot occurrence in advance. In this thesis ANN based prediction engine is used in triggering a proactive DTM mechanism. The benefit of prediction based DTM approach is that it eliminated transient overshoots in temperatures as well as helps us avoid being conservative in setting maximum temperature bounds for the chips. In addition to being predictive, DTM for modern and future multicore chips must combine both core-level as well as interconnect level techniques.

Several researchers are investigating the possibility of designing NoCs with wireless interconnects to reduce the energy consumption in on-chip data transfer. In [10], proposes a wireless NoC based design for the CMOS Ultra-Wideband (UWB) technology. In [24], mm-wave wireless on-chip embedded antennas for intra chip and inter chip communication are designed and evaluated. Possibilities of creating novel architectures aided by the on-chip wireless communication have been explored in [25] and [9]. Design of a wireless NoC using the small-world topology using carbon nanotube (CNT) antennas operating in the THz frequency range is elaborated in [11]. However, integration of these antennas with standard CMOS processes needs to overcome significant challenges whereas mm-wave CMOS transceivers operating in the sub-THz frequency ranges is a more near-term solution. These two works proposed design of hybrid WiNoC architectures using long-range wireless shortcuts. It is shown that WiNoC improves the temperature profile of the NoC switches and links compared to a traditional mesh [26]. In [27], it is demonstrated that incorporating dynamic voltage and frequency scaling (DVFS) in a WiNoC can improve the thermal profile of a multicore chip. In [28], effects of thermal management on wireless NoC architecture were studied. In [29], temperature aware rerouting for wireless network-on-chip is proposed.

However, most of the existing works principally address either thermal management strategies for the processing cores or NoC components individually whereas both contribute to the temperatures of each other significantly affecting overall temperature. Hence, in this work address the local thermal hotspots of multicore chips through a combined dynamic thermal management technique coupling temperature-aware routing strategy and task reallocation.

# Chapter 3. Artificial Neural Network to Predict Temperature profile in a Multi-core chip

The proactive Dynamic Thermal Management (DTM) mechanism in this work, is triggered using Artificial Neural Network (ANN) based temperature predictor. Future temperature is estimated from the performance counters or utilization metrics in advance. This prediction based DTM eliminating the chance of transient overshoot of temperature. As a result, the impact on performance is minimal for these types of predictive thermal estimator [3]. As these thermal estimation schemes need to be implemented on-chip, it is important for these mechanisms to have low computational and area overheads. ANN based prediction mechanism has been shown to perform with high accuracy in various application areas [5]. This work designs a hardware-based ANN thermal predictor that is trained and apply to predict the temperature at any given time based on the utilization of the chip components. An ANN consists of two elements: first element adds products of input and weights coefficients. The second element is a neuron activation function which is a nonlinear function, as explained in section 1.3. The system dependency between input and output can be modeled using the ANN, on training the ANN on a train data set. The training data consists of inputs i.e. utilization of the chip components and their corresponding output i.e. temperature increases due to the utilization of the element. ANN are Multi-Input Multi-Output model that demand hardware overhead. To reduce the hardware overhead and reuse the available resources, the neurons are realized as parallel multiply-accumulate operation (MAC) units.

## 3.1 Creating the training dataset

The desire to have an ANN to model the dependency between utilization of the core and NoC level components with the temperature various components of the chip (cores/switches/links), creates the necessity to generate a training dataset. The utilization of the cores are represented by the percentage utilization of the processors while the utilization of the switches are measured as the ratio of actual buffer occupancy to that of the maximum. The utilization of the links are measures at the attached switches as the ratio of the actual rate of flits transferred over the link to the maximum capacity of the link. All the utilizations are expressed as a percentage. To gather the training data, first random initializations of utilizations of core level and NoC level component is done on a cycle accurate in-house multi-core NoC simulator. The simulator is allowed to run for 3000 cycle with the initialized conditions. At the end of 3000 cycles simulator outputs the temperature change in each cycle. Similarly several random utilization conditions are initialized to the simulator, then the corresponding output temperatures are recorded as the training data. The total training data that was gathered is 3000 x 250 samples. This would help for the prediction of temperature in that particular help in the operating range of the system.

## 3.2 Design of the ANN

The inputs to the ANN are the utilization of all the network element. The output of Base ANN is the temperatures of the cores, switches and links. The Base ANN output is compared to the threshold producing a single bit output. This output signifies if the element has crossed the threshold or not. The trained ANN is a concatenation of three subdivided ANN streams. The subdivided ANN streams are Core streams, Link stream and Switch stream. The inputs to all the streams remain the same, i.e. the utilization of all the network elements and time to be predicted

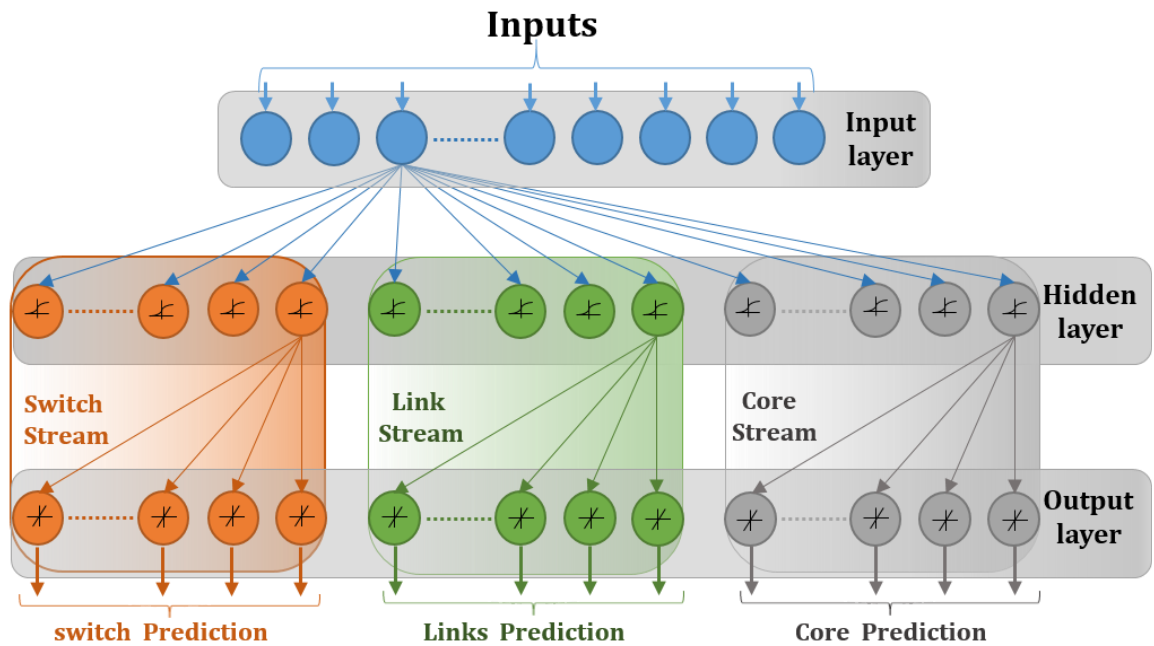

Fig. 3: Trained ANN structure.

This structure of ANN results in better accuracy and decreases the number of connections between hidden layer and output layer i.e. the fully connected neurons exists only between the hidden neurons of cores and output neurons of the cores. The reduced number of connections between the hidden neurons and the output neurons decreases the latency of prediction. . The nntool box of Matlab is used to train to neural network. The above Fig.3 shows the trained ANN structure. The number of hidden layer neurons used for the Cores stream is 250 whereas for the Switches and Links stream are 50 and 100 respectively. These hidden layer neurons were selected by a number of trials, resulting for the best accuracy and low latency. There were a number of experiments done using different kinds of activations function but the best accuracy was produced by using the sigmoid activation function for hidden layer and linear activation function for the output layer. To improve the latency of calculation, the number of neurons were decreased from 700 to 400. It was observed that there is negligible loss on accuracy for change in the number of neurons.

### 3.3 ANN hardware

To have an efficient ANN realization with the use of available resource in a core, the neurons are realized as multiplier and accumulator (MAC) units and work in parallel. Fig.4 shows the designed ANN hardware. The computation to predict the temperature, starts no sooner the utilization information packets are received at the ANN core in the pipelined fashion. This utilization values are represented as 8 bit numbers. We propose to use wireless interconnections for the transfer of information to and from the ANN to the different components of the chip. The components packetize their current utilization values and send it to the ANN via the nearest NoC switch. The only wireless switch with the token to pass network utilization can send the information. There are 240 REG arrays each of size 8 bit.

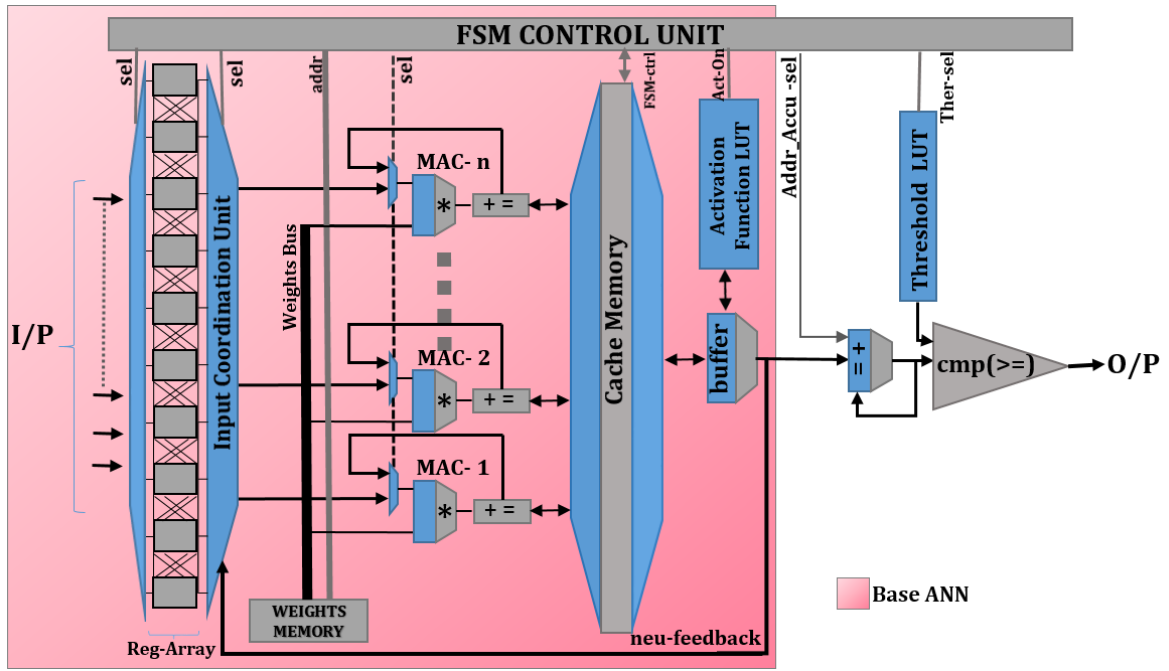

Fig. 4: ANN hardware

The number of neuron operations that are occurring in parallel are equal to the number of available MACs. As the number of MAC units that work in parallel are less than the number of inputs, we use register arrays to store the inputs. Inputs are being multiplied with the corresponding weight values that are fetched from the weight memory. The. FSM control unit synchronizes the received inputs, weights and directing these values to the right MAC unit. MAC outputs are stored in the cache. When all the connections to a particular neuron are complete then the output is passed through the activation function via buffer. The output of the LUT is feedback to the input coordination unit and REG array, only if the computation was done between first two layers (input layer and hidden layer). The process is repeated for the hidden layer and the output layer. For the first iteration the predicted increase in temperature is added to the initial temperature of the chip and then accumulated. From the second iteration onwards predicted temperature is added to the accumulated result. The accumulator result is passed through a comparator, where it is compared with the threshold. The comparator outputs a single binary bit. This bit signifies that, the corresponding network element is in the 'ON' state and the zero bit signifies the 'OFF' state of the element. The comparator output is broadcasted wirelessly, which is used by network elements for efficient Distance Vector Routing (DVR) and Task Rerouting (TR). The ANN predicts the future temperature for a given utilization every 100,000 cycles. Hence, the bandwidth demand of these DTM related packets on the wireless interconnection is 0.054Gbps. This being a small fraction of the wireless bandwidth of 16Gbps does not significantly affect the performance. For efficient DVR and TR to be performed that results in a chip temperature less than the threshold, the 240 status bits are to be transmitted within 1,000,000 clock cycles.

## 3.4 Evaluation of the proposed thermal ANN predictor

To evaluate the hardware overheads of the proposed ANN based thermal estimator, the ANN structure is implemented in RTL and synthesized it using 65nm standard cell libraries from CMP. The total area and power of 10 MAC units is found to be 1.82572 mm2 and 501.11$\mu W$ respectively. The total delay in the circuit for the overall ANN architecture is 1.1226$\mu s$. We use 240 register arrays each of size 8 bits, as the register array is used to store the inputs. The activation function and temperature threshold value is stored on two separate LUTs. Table I shows the memory requirements for different elements of the ANN structure. In addition to the thermal estimator, the wireless transceivers required for the exchange for the thermal control flits and onchip data also require area overheads as explained next.

| Device | Memory |
|---|---|
| Total Input Reg Bank | 1.2KB |
| Weight Memory | 300KB |
| Threshold LUT | 0.048KB |
| Activation Function LUT | 1.32KB |
| Total memory required | 302.568KB |

Table I: Memory requirement for the ANN based thermal estimator

Table II shows the root mean square errors of the ANN and LUT-based thermal predictors with respect to the temperature traces generated using HotSpot. We have considered HotSpot as our baseline as it is a very accurate thermal modeling tool with error less than 1% [42]. A full system simulator GEM5 [35] was used to obtain the detailed processor and network-level information on SPLASH-2 [36] and PARSEC [[37] benchmarks. We consider a system of 64 alpha cores running Linux within the GEM5 platform for all experiments. The memory system is MOESI_CMP_directory, setup with private 64KB L1 instruction and data caches and a shared 64MB (1MB distributed per core) L2 cache. The processor-level utilization statistics generated by the GEM5 simulations are incorporated into McPAT simulator [38] to determine the processor-level power statistics.

| Benchmark | Root mean square error (ºC) | |
|---|---|---|
| | ANN based thermal predictor | Event-driven LUT based thermal predictor |
| CANNEAL | 2.9026 | 1.2773 |
| BODYTRACK | 2.8830 | 1.2216 |
| VIPS | 2.8009 | 1.1209 |
| FLUIDANIMATE | 2.8374 | 1.0822 |
| SWAPTION | 2.7784 | 1.1428 |
| FREQMINE | 2.7984 | 1.1300 |
| FFT | 2.7942 | 1.1754 |
| RADIX | 2.8593 | 1.0894 |
| LU | 2.8465 | 1.2317 |

Table II : Root Mean Square Error of the ANN based predictor and event driven LUT based thermal estimator with respect to HotSpot

From table II, it can be seen that root mean square error for the proposed ANN based thermal predictor varies from 2.7ºC~2.9ºC whereas for the event driven LUT based predictor with fine-grained time interval (100μs), it is between 1.08ºC~1.23ºC.

The ANN based prediction clearly has a greater mean square error when compared to the LUT method. This is because it is practically impossible to train the ANN for all possible input combinations. Running the ANN for input combinations that it has not yet been trained for will result in some error in its prediction. However, as the ANN has the capability to learn on its own as we keep using the ANN for thermal estimation, the mean square error will continue to reduce below the values specified in Table II.

| Time interval | 100μs | 1ms | 10ms |
| --- | --- | --- | --- |
| **Number of rows required to reach steady state** | ~3000 | ~1700 | ~1000 |
| **Required memory** | 1382MB | 784MB | 460MB |

Table III: Memory requirement for LUT based thermal estimator

Although the LUT based thermal estimation method is more accurate it requires an impractical amount of memory storage due to large number of temperature samples it has to store. Table III shows the memory requirement for the event driven LUT based thermal estimator assuming 8 bit encoding to store the temperature values. Three different non-uniform time intervals is used to calculate the size of the LUT. This amounts to a memory requirement of 1382MB, 784MB and 460MB for time intervals of 100μs, 1ms, and 10ms respectively. Such a large memory requirement is neither practical nor scalable. On the other hand, the overheads of the ANN based predictor is negligible in comparison to the LUT. Hence, we adopt the ANN-based temperature predictor for our thermal management scheme.

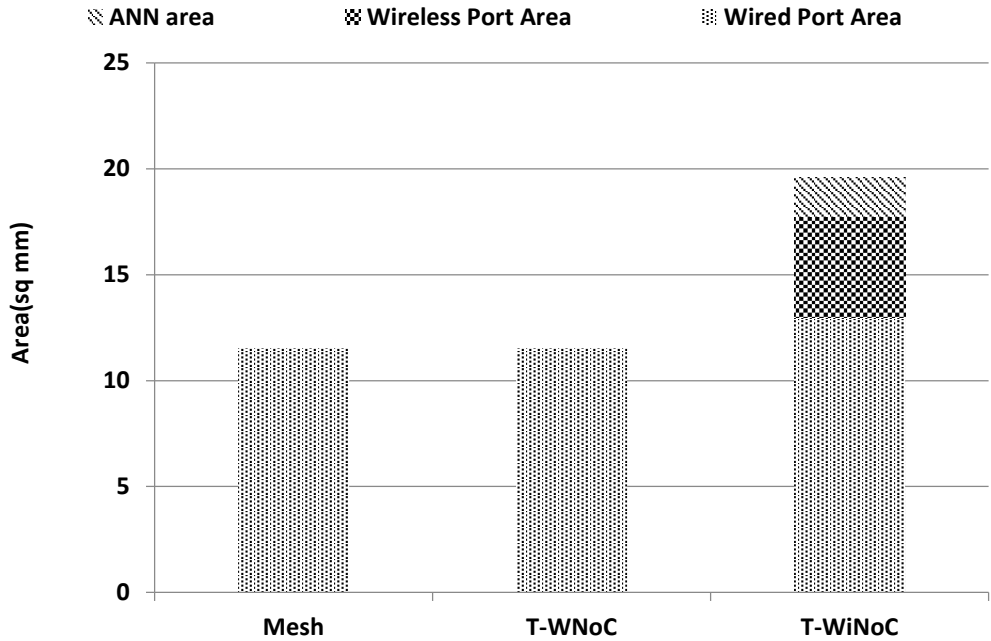

Fig. 5: Area overheads of different architectures

The area of a NoC switch depends on the number of ports, their virtual channels and buffer depths. On the other hand, the area of the wireless port is the total area of the transceiver circuit, antenna and the area required to buffer the wireless flits. The wireless transceiver circuits require an area of 0.3mm$^2$ each [4] and the area of the antennas is 0.33mm$^2$ each. The total area for the different NoC architectures considered in this paper is shown in Fig. 18. The T-WNoC architecture contains same number of wired ports as planar mesh as they have exactly same number of links. In case of T-WiNoC, each electronic switch has an additional VC to send utilization information efficiently. Besides that, additional area overhead of T-WiNoC architecture compared to mesh and T-WNoC is due to the area of wireless transceivers, antennas, the area of the ports associated with those as well as the area of the ANN. The total area overhead of implementing T-WiNoC architecture along with the ANN is 19.58 mm2, which is 4.9% of the total die area of 400 mm2.

# Chapter 4. Integration of ANN based Thermal predictor with Dynamic Thermal Management

The efficiency of a Dynamic Thermal Management (DTM) scheme depends on proper thermal estimation and response delay of the control mechanism. From this viewpoint, DTM methods can be divided into two types: reactive DTM and proactive DTM. In case of reactive DTM methods, generally on-chip thermal sensor is used to sense the temperature. This allows the hardware to execute at full speed and initiate a corrective measure only when the temperature reaches a thermal limit and invokes the temperature control mechanism. However, effectiveness of such system relies on associated response delay. Due to the response delay, thermal thresholds need to be set conservatively to avoid temperature overshoot that in turn impacts system performance. Moreover, on-chip thermal sensors are sensitive to process variations and without recalibration, can report error temperatures.

In proactive or predictive DTM method, future temperature is estimated from the performance counters or utilization metrics in advance, thus eliminating the chance of temperature overshoot. As a result, the impact on performance is minimal for these types of predictive thermal estimator. However, as these thermal estimation schemes need to be implemented on-chip, it is important for these mechanisms to have low computational and area overheads. Artificial Neural network (ANN) based prediction mechanism has been shown to perform with high accuracy in several application areas. In this work, a hardware-based ANN-based thermal predictor designed, that was trained and applied to predict the temperature at any given time based on the utilization of the chip components. The design of the ANN is explained in Chapter 3 of this thesis.

In the wireless NoC, data is transferred via wormhole routing using virtual channel (VC) based switches. Data packets are broken down and transferred in the form of flow control units or flits which are the smallest amount of information in a packet that can be transferred between adjacent switches in one clock cycle. In addition to normal VCs to transfer data flits the switches will have one reserved VC to send the utilization and routing control packets. All the switches will send link and switch packetized their activity information to their nearest WIs that in turn will send these packets to the scheduler. Based on this information, scheduler will predict the temperature using the thermal predictor to trigger the DTM scheme.

## 4.1 Combined Dynamic Thermal Management Scheme Coupling Task Reallocation and Rerouting

A temperature-aware task reallocation with network level rerouting was proposed for equipping the scheduler in the multicore chip with a combined pro-active thermal management technique. This proposed technique improves the thermal profile while minimizing the impact on system performance. Artificial Neural Network is used to predict the temperature and take corrective measures instead of the reacting to the temperatures measured from on-chip sensors.

Redistributing the workload, temperature-aware task reallocation will reduce the temperature of the cores. Temperature-aware task reallocation may not reduce the network hotspots. A dynamic routing approach taking into account switch and links temperatures was proposed and investigated. The idea is to dynamically configure routing paths in response to temperature increases so that heat dissipation is distributed better and points of failure are avoided in the NoC. One target temperature, $T_{th}$ for activating the combined dynamic thermal management scheme was considered. Task reallocation essentially changes the traffic among cores by redistributing tasks, triggering task-reallocation may affect rerouting decisions. For this reason, a sliding window method to trigger the combined scheme to avoid oscillation between task-reallocation and rerouting was employed. The ANN based thermal predictor described in Chapter 3, predicts component temperatures starting from next time instance to all time instances within the window. The scheduler using the combined DTM scheme activates a temperature-aware task reallocation strategy, if the predicted temperature of any core increases beyond $T_{th}$. Now, if predicted switch or link temperatures is above $T_{th}$, then the scheduler slides the window to predict core temperature within next window interval, and if any of the cores crosses $T_{th}$ on the next window, scheduler triggers task allocation instead of rerouting, otherwise, the scheduler triggers temperature-aware rerouting. The scheduler with the DTM module and ANN is housed, hence it is considered to be in one of the cores of the chip. The combined thermal management scheme with the sliding window technique is schematically shown in Fig.6. In our implementation, the neural network based thermal predictor will periodically predict the temperature of all the chip components. If any component temperature as predicted by the ANN exceeds the target temperature, that component is flagged and a notification flit containing either distance vector or DTM triggering information is sent to the WI nearest to that particular component. From the WI this thermal control message of one flit is routed downstream to the target switch whose associated component has crossed the target temperature. The format of the thermal control message is shown in Fig.7.

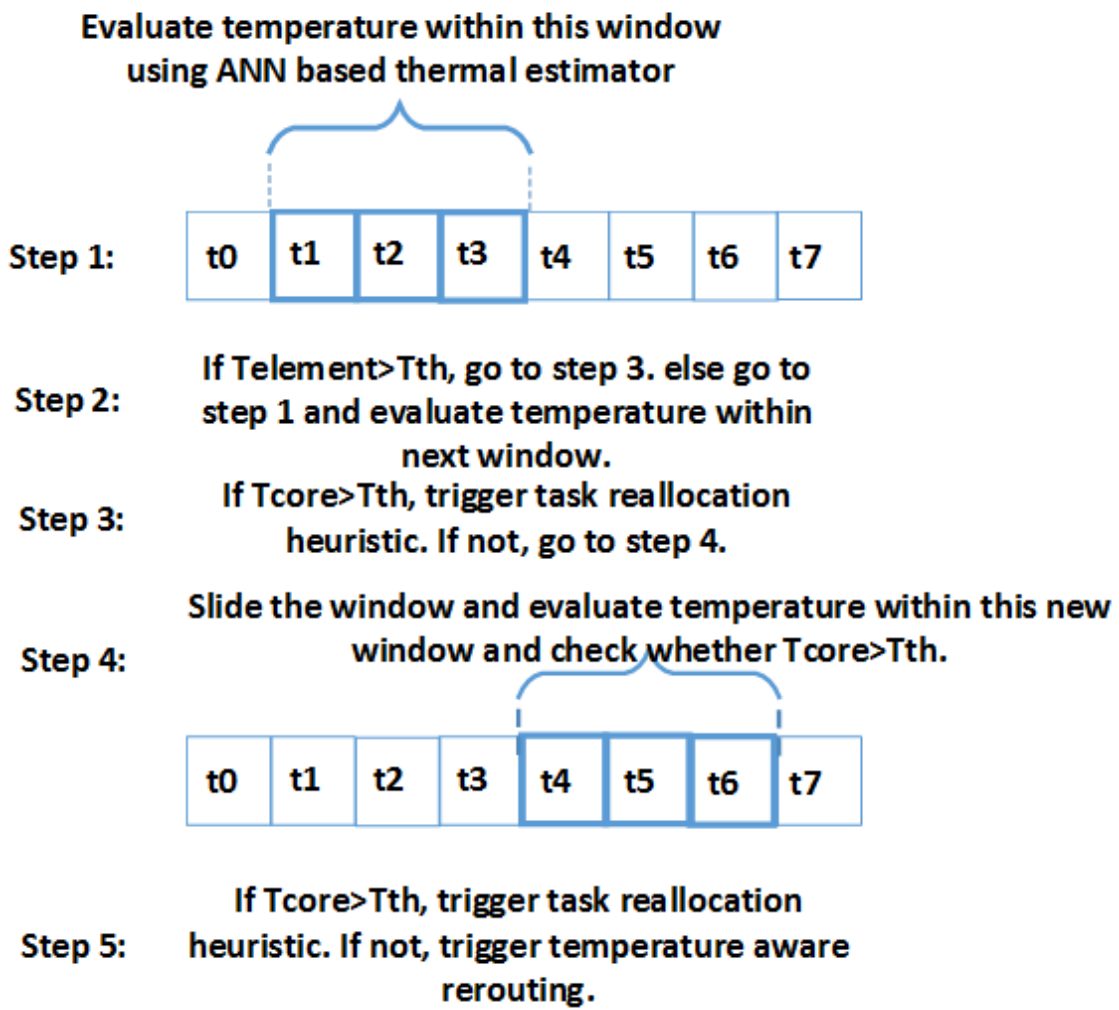

Fig. 6: combined dynamic thermal management scheme

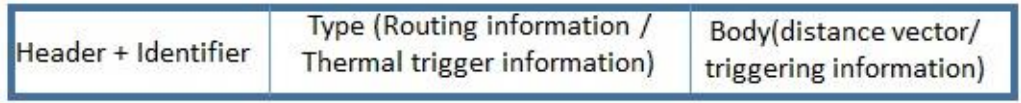

Fig. 7: Thermal control message format

A task reallocation heuristic that considers both temperature characteristics of the chip as well as the performance of the WiNoC is used. The Artificial neural network based thermal predictor is used to trigger the task reallocation. To evaluate resultant temperature, performance, and energy-efficiency of data communication over the T-WiNoC architecture, task reallocation heuristic with temperature aware rerouting is combined. A novel task reallocation algorithm based on future temperature trends (FTT) was proposed in [4]. It is a temperature-prioritized method where task reallocation is done such that threads with highest power consumption are allocated to either the fastest cooling or slowest heating cores. We adopt and modify this algorithm to factor in the performance of the NoC as well as temperature of the links and switches along with the temperature of the cores.

Taking into account switch and link temperatures, task reallocation is also combined with dynamic temperature aware routing approach. The temperature-aware rerouting scheme is based on adapting the Distance Vector Routing (DVR) algorithm [32] based on the Bellman-Ford equation for the NoC environment. DVR was designed to support routing under dynamic conditions in large scale networks. It is the de-facto standard for intra-domain routing over the internet where varying congestion conditions and traffic flows affect network integrity [32]. Neighboring switches maintain the cumulative path cost to all other nodes in the network also known as the distance vector. In addition each switch also has a forwarding-table containing the information about the next hop for all destinations. If a change in the link cost is detected, the distance vectors are packetized along with the time stamp and advertised between the nearest neighbors by all the switches. The routing table of a switch is updated every time a change in the link cost is triggered by the temperature of a NoC component exceeding the threshold. Thus, the routing tables in the switches may change several times until the entire network converges. This may result in deadlocks. In order to avoid such deadlocks, the proposed rerouting scheme uses two routing tables for every switch. The old routing table is used to route all the data packets until the network converges. Only after network convergence the newly calculated routing table is used to route all the newly generated data packets. The worst convergence delay of the routing protocols are determined by the time necessary for propagating new paths. This propagation delay depends on the connectivity of the network i.e. maximum diameter of the network and message processing delay of the switches [33]. In our implementation, each switch has one reserved Virtual Channel (VC) to process the packetized distance vector information. Moreover, these control packets are of short length. As a result, the message processing delay is reduced significantly. Besides, these control packets are using wireless links that are placed to optimize the network performance by minimizing the average hop count. The well-connected smallworld network augmented with single hop wireless links ensures faster convergence of the DVR algorithm. In practical scenarios, the network paths using DVR eventually converge to the shortest path routing tree obtained through Dijkstra's algorithm [32]. The old forwarding table is used till the entire network reaches convergence to avoid multiple paths between same source/destination pairs. Deadlock is avoided as at any point of time the flits are transferred over paths along the shortest path routing tree. Experimentally, it is found that a period of 600 cycles is sufficient for the entire network to converge in a 64 core system. Hence, all switches start using the new routing tables after 600 cycles after the rerouting is triggered. Switches know the exact time to start as the ANN broadcasts the time stamp of the last prediction when it sends the control flit that triggers the rerouting.

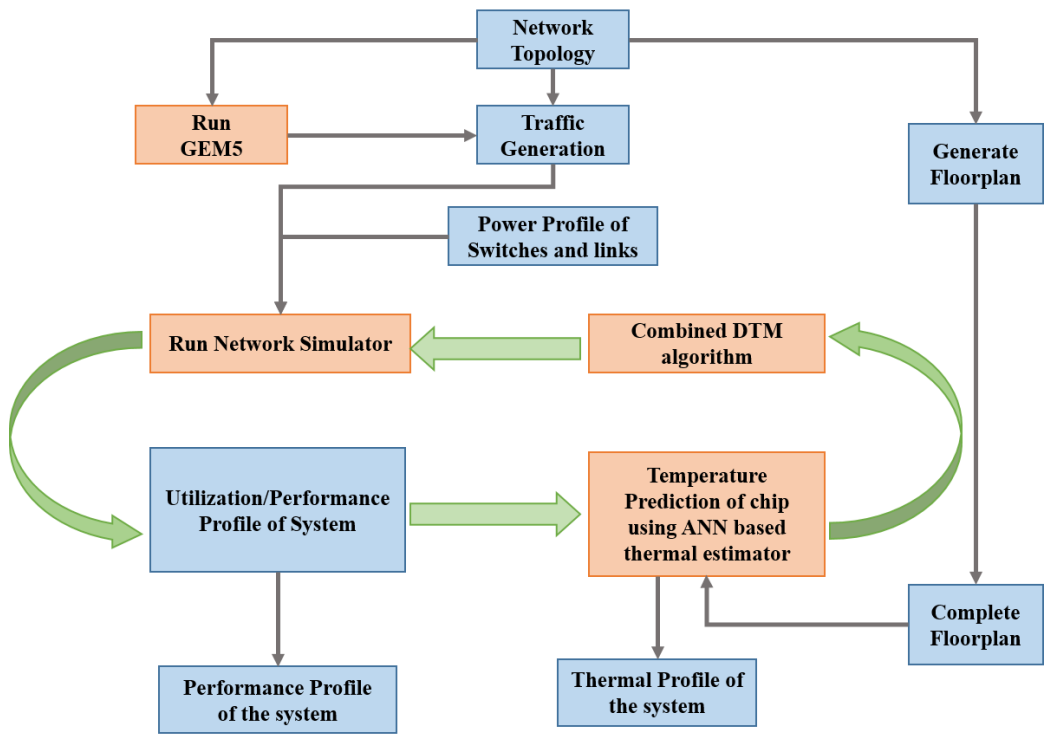

Fig. 8: Thermal profile evaluation simulation flow

## 4.2 Thermal Characteristics with combined DTM and ANN

This section evaluates the performance and temperature profile of a multicore chip with 64 cores equipped with the proposed combined dynamic thermal management scheme. To evaluate the temperature profiles, performance and energy consumption of the NoCs an integrated simulation environment as shown in Fig. 8 was developed. The detailed processor and network-level information on SPLASH-2 [36] and PARSEC [[37] benchmarks was obtained using a full system simulator GEM5 [35]. A system of 64 alpha cores running Linux within the GEM5 platform was considered for all experiments. The memory system is MOESI_CMP_directory, setup with private 64KB L1 instruction and data caches and a shared 64MB (1MB distributed per core) L2 cache. The processor-level utilization statistics generated by the GEM5 simulations are incorporated into McPAT simulator [38] to determine the processor-level power statistics. The traffic interaction patterns for each benchmark obtained from Gem5 are used in the NoC simulator to obtain the NoC performance in terms of peak bandwidth, average network latency, average packet energy and the instantaneous utilization and power of the NoC components. Based on the temperature profile obtained from the ANN thermal estimator and the effect of the combined DTM scheme, routing paths and task mappings are updated dynamically in the NoC simulator to capture the transient effects. The NoC architecture is characterized using a cycle accurate simulator that models the progress of the data flits accurately per clock cycle accounting for those flits that reach the destination as well as those that are stalled. As is common in NoCs with wormhole switching, we do not allow packets or flits to be dropped in our architecture but rather stall the communication pipeline in case of congestion. The width of all wired links is considered to be same as the flit size, which is considered to be 32 bits in this paper. We consider a moderate packet size of 64 flits for all our experiments. Similar to the wired links, we have adopted wormhole routing in the wireless links too. The particular NoC switch architecture adopted in this work has three functional stages, namely, input arbitration, routing/switch traversal, and output arbitration [39]. Each switch port has four virtual channels for data transfer and one reserved virtual channel for control packet each with a buffer depth of 2 flits. The wireless ports have an increased buffer depth of 8 flits to avoid packet dropping while waiting for the token. Increasing the buffer depth beyond this limit does not produce any further performance improvement for this particular packet size, but will give rise to additional area overhead [9]. The switches are synthesized from RTL level designs using 65nm standard cell libraries from CMP [40], using Synopsys. The delay and energy dissipation of these digital components are obtained from the post-synthesis RTL models. The NoC switches are driven with a clock of frequency 2.5 GHz at 1V. The delays and energy dissipation on the wired links were obtained through Cadence simulations taking into account the specific lengths of each link based on the established connections in the 20mmx20mm die following the topology of the NoCs.

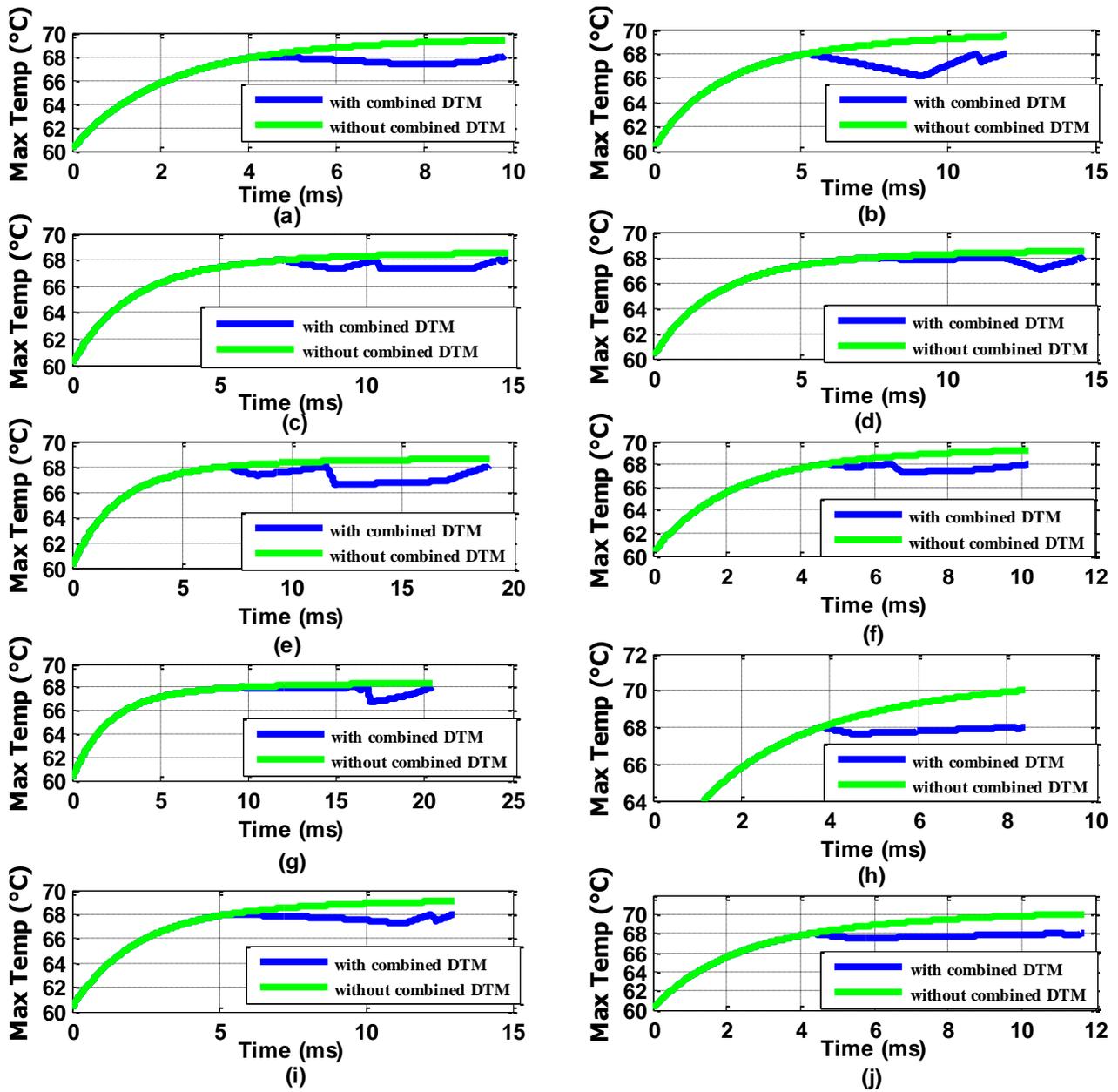

Fig. 9: Maximum chip temperature with and without combined DTM for (a) CANNEAL, (b) BODYTRACK, (c) VIPS, (d) DEDUP, (e) FLUIDANIMATE, (f) SWAPTION, (g) FREQMINE, (h) FFT, (i) RADIX, and (j) LU traffic.

The wireless transceiver adopted from [9] is designed and characterized using the TSMC 65-nm CMOS process. This design is shown to dissipate 36.7mW for long range on-chip communication distances of the order of 20mm while sustaining a data rate of 16Gbps with a bit-error rate (BER) of less than $10^{-15}$.

The thermal profile is evaluated for the multicore chip with the T-WiNoC equipped with the combined DTM mechanism in presence of application based workloads. To evaluate the thermal characteristics of the combined DTM at high target temperature, the chip is first warmed up to 60°C. The initial task allocation is considered to be random. Although we can set any arbitrary temperature higher than ambient as target temperature, for the combined DTM, a target of 68°C is set at the beginning of the simulation for this experiment. Fig. 9 (a), (b), (c), (d), (e), (f), (g), (h), (i), and (j) show the peak chip temperature for CANNEAL, BODYTRACK, VIPS, DEDUP, FLUIDANIMATE, SWAPTION, FREQMINE, FFT, RADIX, and LU traffic respectively.

It can be observed from Fig. 9 that peak chip temperature without DTM trends to increase in an exponential manner while with proposed combined DTM scheme, after reaching the target temperature, temperature aware re-routing or task reallocation is triggered depending on whether NoC components or cores reach the target temperature first. In all cases, peak chip temperature stays below the target threshold. In case of CANNEAL, BADYTRACK, SWAPTION, and RADIX traffics, the effect of combined DTM is more prominent. This is because in these benchmarks, few cores have very high activity and frequency of communication, while others have relatively uniform communication pattern among themselves. Thus, temperature of the NoC components connecting those cores or the core itself can get very high compare to others creating a hotspot like behavior. As a result, redistributing traffic through cooler links/switches by bypassing hot links/switches or remapping tasks to cooler cores can obtain more uniformly distributed temperature profile on the chip.

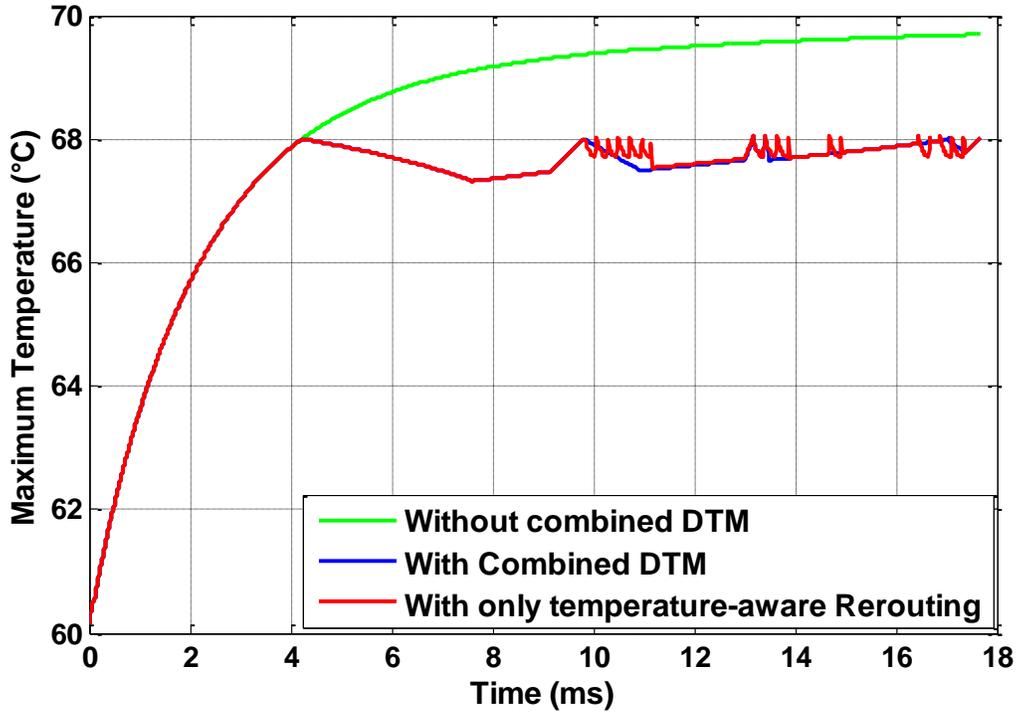

Fig. 10: Maximum chip temperature with and without combined DTM scheme for CANNEAL traffic running for long duration

The effectiveness of the combined DTM is more evident from Fig. 10 where long-term transient temperature response of two systems: system with combined DTM and system with only temperature-aware rerouting are shown in presence of CANNEAL traffic. Compared to a system where combined DTM is used as thermal management policy, in a system with only temperature-aware rerouting as in [29], DTM triggers more frequently. This results in frequent oscillations in routing paths.

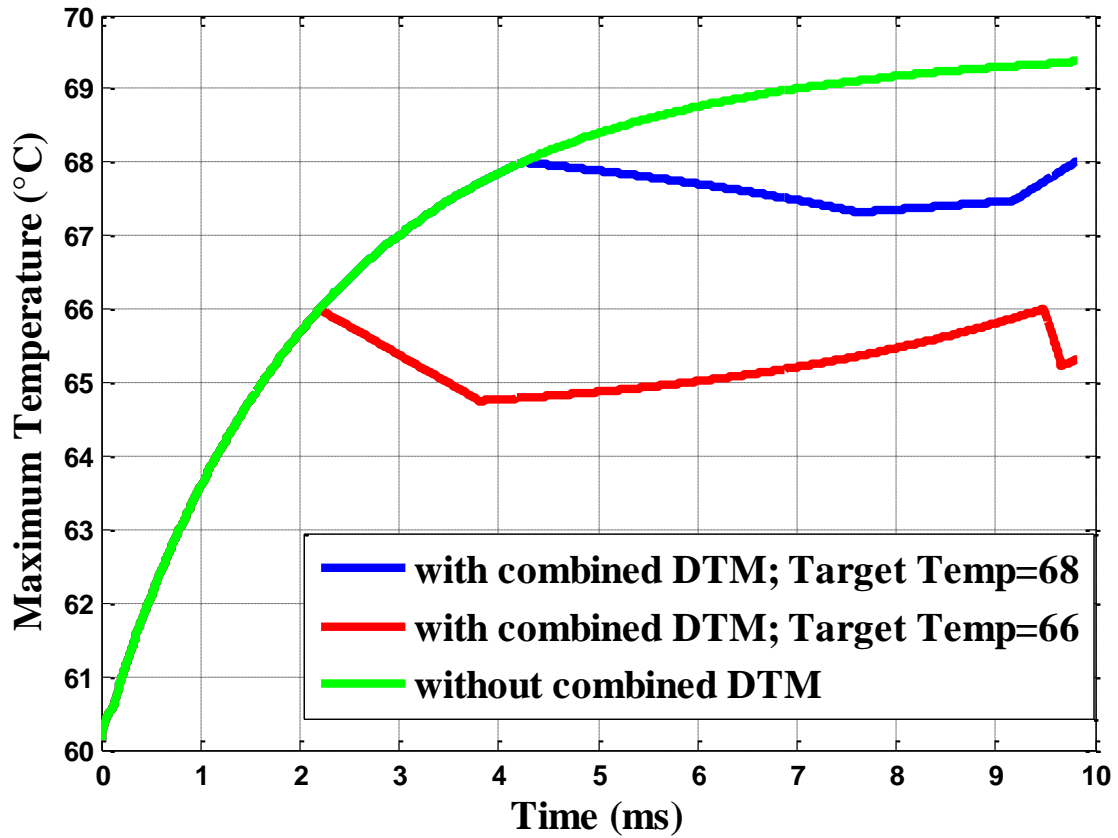

Fig. 11: Transient temperature response for CANNEAL with two different target temperature

This is because of the absence of core-level DTM policy. As time progresses, temperature of the cores also start to increase which in turn affect the temperature of links/switches connected with those cores. As a result, temperature of many links/switches increases relatively quickly making it difficult to find alternative paths to avoid these hot components, which in turn results in oscillation in routing paths.

Fig. 11 shows transient temperature response for CANNEAL traffic with two different temperature thresholds. From Fig. 11, it can be seen that combined DTM scheme can control the maximum chip temperature for both target thresholds. However, compared to higher temperature threshold, the system with lesser temperature threshold triggers combined DTM more frequently that in turn can affect the performance. We will discuss the performance tradeoff of implementing combined DTM scheme in the next section.

# Chapter 5. Conclusion

Thermal concerns in a multicore chip are aggravated by aggressive scaling of system size and inter-core traffic interaction over the NoC fabric. A temperature-aware predictive dynamic thermal management technique combining both task reallocation and rerouting techniques designed for the WiNoC architecture successfully restricts the temperature of the NoC components near a target threshold value. The wireless interconnection with broadcast abilities along with the prediction capability of the ANN improve the reaction time of multicore system ensuring on-chip temperatures never exceed the target threshold. Such a system can be used to design thermally efficient multicore chips with wireless NoC fabrics with minimal overhead. This design is particularly suitable for applications where the utilization of the cores and NoC components are heterogeneous where some parts are utilized to a large extent creating local thermal hotspots leaving other components relatively cooler. Under these circumstances the workload can be redistributed to relatively cooler parts dynamically to ensure a more homogeneous thermal profile.

This thesis proposes an ANN based prediction engine to predict the temperature profile of the chip component. Combining three subdivided ANN streams, the proposed ANN structure shows root mean square error of 2.7ºC~2.9ºC with respect to the HotSpot. It is possible to further improve the accuracy of the ANN by increasing the number of hidden layer neurons. However, considering the hardware implementation, 240 hidden layer neurons were used to provide best tradeoffs in overhead and accuracy.

Compared to the LUT based thermal estimator, the ANN requires a total memory of 302.568KB, which is 0.022% of that of a LUT based thermal estimator. It is seen that bandwidth demand of the DTM related packets on the wireless interconnection is 0.054Gbps, which is a small fraction of the overall wireless bandwidth of 16Gbps. The total area and power of 10 MAC units is found to be 1.82572 mm$^2$ and 501.11µW respectively. The total delay in the circuit for the overall ANN architecture is 1.1226µs. Increasing the number of parallel MAC units can decrease the computational time with the tradeoff for ANN hardware area.

The efficiency of the ANN hardware can be improved by using state-of-the art low-powered circuit design techniques like neuromorphic engineering techniques. However, it requires analog circuitry for the ANN structure. Due to the learning ability of the ANN systems, it is possible to design future generations of self-learning and self-adjusting computer chips. To satisfy the future computational needs, it is expected to have 100x more cores than the current state of the multicore design. To predict the temperature profile of such complex system, complex brain inspired algorithms like the convolutional neural network can be used. The ANN can also be used to solve the onchip prediction problems that cannot be modeled using a polynomial equation.